\documentstyle[11pt,epsfig]{article}

\begin{document}
\title{Initial Conditions and 
the Structure of the Singularity in Pre-Big-Bang Cosmology}
\author{A. Feinstein$^{a,}$\footnote{{\tt wtpfexxa@lg.ehu.es}}\hspace*{0.1cm}, 
K.E. Kunze$^{b,}$\footnote{{\tt kunze@amorgos.unige.ch}}\hspace*{0.1cm} and
M.A. V\'{a}zquez-Mozo$^{c,d,}$\footnote{{\tt vazquez@wins.uva.nl, M.Vazquez-Mozo@phys.uu.nl}}
\\ \\
{\normalsize\sl $^{a}$ Departamento de F\'{\i}sica Te\'orica, 
Universidad del Pa\'{\i}s Vasco,}\\ {\normalsize\sl  Apdo. 644, E-48080 Bilbao, Spain} \\
{\normalsize\sl $^{b}$ D\'epartement de Physique Th\'eorique, Universit\'e de Gen\`eve,} \\
{\normalsize\sl 24 Quai Ernest Ansermet, 1211 Gen\`eve 4, Switzerland }\\
{\normalsize \sl $^{c}$ Instituut voor Theoretische Fysica, Universiteit van Amsterdam,
}\\ {\normalsize\sl Valckenierstraat 65, 1018 XE Amsterdam, The Netherlands }\\
{\normalsize\sl $^{d}$ Spinoza Instituut, Universiteit Utrecht, Leuvenlaan 4,}\\
{\normalsize\sl  3584 Utrecht, The Netherlands}}

\date{}

\maketitle

\abstract{We propose a picture, within the pre-big-bang approach, in which 
the universe emerges from  a bath of {\em plane} gravitational and dilatonic waves.
The waves interact gravitationally breaking the exact plane
symmetry and lead generically to gravitational collapse resulting in a
 singularity with the Kasner-like structure. The analytic relations 
between the Kasner exponents and the initial data are explicitly evaluated
and it is shown that pre-big-bang inflation may occur within a 
dense set of initial data.
Finally, we argue that plane waves carry zero gravitational entropy and thus
are, from a thermodynamical point of view, good candidates for the universe 
to emerge from.}

\section{Introduction}

The low energy effective equations of string theory provide
cosmological solutions which might be applicable just below the 
string scale in the very early universe.
In the pre-big-bang (PBB) scenario, suggested naturally by the spirit and the 
symmetries of Superstring theory,
the universe
starts in a low curvature, low coupling regime and then enters a stage
of dilaton driven kinetic inflation \cite{gasp}.  To address one of the
main problems of cosmology, namely the problem of the initial conditions, 
this interesting picture has been  developed further in \cite{buon},
where the authors suggest that  the initial state of the universe
could have  consisted of a bath of
gravitational and dilatonic waves, some of which would have  collapsed leading  to
the birth of a baby inflationary universe. These PBB bubble universes 
would give rise, finally,
after a yet-to-be clarified graceful exit mechanism, to the observed
Friedman-Robertson-Walker (FRW) world.

The main purpose of this paper is to 
develop a modified {\em realization} of the PBB bubble picture
of Buonanno,  Damour and Veneziano \cite{buon}, 
in which the spherically symmetric collapse 
 leading to inflationary PBB solutions
is substituted by the interaction of strictly plane waves. This modification 
affects only the initial state of the universe, while
near the (spacelike) caustic singularity the model shows similar behaviour to
that
discussed in \cite{buon}, leading  to Kasner-like structure.
Representing exact solutions to the classical string equations of motion to all
orders in the inverse string tension \cite{horstief}, it looks rather natural
to modify the PBB picture by incorporating plane waves into the postulate of
``asymptotic past triviality" \cite{buon}. Moreover, this picture
is  attractive not only due to the exactness of plane wave
backgrounds for string propagation, but  most importantly, because 
of mutual ``fatal attraction" exercised by the plane waves which leads to an 
inevitable  gravitational collapse independently of their
strength, unlike in the spherical picture.

In the scenario we propose, one starts with a model universe in a low coupling, 
low curvature
regime   with plane gravitational and matter waves which eventually 
will gravitationally interact \footnote{In the gravitational 
sector we limit
ourselves to the discussion of constantly polarized waves (diagonal metrics) 
on the
grounds that in string theory the collision of gravitational waves with 
variable polarization may be generically mapped via T-duality into another 
problem where the variable polarization of the incoming waves is transformed 
into an non-vanishing 
value of the $B_{\mu\nu}$ field \cite{flvm}.}.
Colliding plane wave space-times have been investigated in detail
in general relativity (see \cite{grif} and references therein). A generic 
feature of 
the interaction of two plane waves is the formation of a strong space-like 
curvature 
singularity in the future \cite{tip}.
In the context of the PBB scenario, this singularity
can be re-interpreted as  an ordinary cosmological singularity. 
 
The approach to the singularity from the past occurs through a Kasner-like 
behaviour, and this, in  turn, can be analytically related to the
initial data. This will provide the framework for a future quantitative study 
of the important problem of fine tuning of initial conditions
\cite{turn,jrd,kl,fvm} leading to {\em inflationary} behaviour as
$t\rightarrow 0^{-}$. This problem
is well defined in our picture, unlike  in the 
 spherically symmetric case where a similar analysis does not seem to be 
possible. In the case at hand, the problem is a direct 
generalisation of the problem 
of determining Kasner exponents in scattering of pure plane gravitational
waves with constant
polarization \cite{yurt}.
As far as the technical part of this paper is 
concerned, some already  well-known
results from general relativity  and mathematical cosmology will be used and re-interpreted 
in a new light.

The whole question of naturalness of initial conditions
is far from being settled and is under current discussion in the literature.
The authors of \cite{kl} have found that the solution of the flatness problem 
requires the introduction of two huge dimensionless parameters. On the other
hand in \cite{gas} it was argued that this might not constitute such a fine
tuning if the initial scale is taken to be the whole horizon and not just
the Planck scale. From a different point of view the genericity of PBB
cosmology has been addressed in \cite{jrd,gr}, showing that plane waves might be 
considered a generic initial state of PBB cosmology. In this paper, however, we will not
enter into the discussion of whether fine tuning is needed in order to solve the
flatness problem, but will rather stick to the possibility of inflation, leaving the
resolution of these other questions to future work.

The paper is organized as follows. In Sec. 2 we review briefly the initial value problem 
for the collision of dilatonic and gravitational waves. In particular we  provide
a closed expression relating the Kasner exponents which characterize the asymptotic 
geometry near the caustic singularity with the initial conditions for the metric functions
and the dilaton. In Sec. 3 we  apply these results to investigate the range of
initial conditions leading to PBB inflation, and whether these conditions are
naturally met in the collision of plane gravitational waves. 
In Sec. 4 we
study two particular geometries in the interaction region, the Nappi-Witten solution 
\cite{nappi},
and a family of Kantowski-Sachs metrics studied in \cite{fvm}. 
Finally, in Sec. 5 we will use thermodynamical considerations to argue that 
plane waves are good candidates to represent the primordial PBB universe, 
summarize our conclusions and indicate possible future directions of research.

\section{Colliding plane waves with aligned polarization}

One of the basic assumptions of the PBB scenario is the so-called ``asymptotic past
triviality" (APT) hypothesis. According to it, the universe starts in the asymptotic past
in  a low curvature and low string coupling  regime where the physics is accurately
described in terms of tree level string theory. The effective dynamics of the 
long-range fields is thus governed by the leading terms of the 
effective low energy string action where both quantum and $\alpha'$ corrections 
are ignored. In four dimensions this action is given in the string frame by \cite{cfmp,r4}
\begin{equation}
S=\int d^4 x\sqrt{-g}e^{-\phi }\left(R+g^{\alpha \beta }\partial _\alpha \phi
\partial _\beta \phi -\frac 1{12}H^{\alpha \beta \gamma }H_{\alpha \beta
\gamma }\right)\label{act}
\end{equation}
where the dilaton $\phi$ and the antisymmetric tensor field strength $H_{\alpha \beta \gamma
}=\partial _{[\alpha }B_{\beta \gamma ]} \label{H}$ are introduced.
Furthermore, we will assume throughout that the extra six spatial dimensions
are compactified in some internal appropriate manifold  considered 
to be non-dynamical.

Applying the conformal transformation
\begin{equation}
g_{\alpha \beta }\rightarrow e^{-\phi }g_{\alpha \beta }.
\end{equation}
the action can be written in the usual Einstein-Hilbert form (Einstein frame).
In this frame the equations of motion are given by \cite{r4} 
\begin{eqnarray}
{R}_{\mu\nu}-\frac{1}{2}{g}_{\mu\nu}{R}
&=&^{(\phi)}{T}_{\mu\nu}+ ^{(H)}{T}_{\mu\nu}
\\
{\nabla}_{\mu}\left[\exp(-2\phi){H}^{\mu\nu\lambda}
\right]&=&0\label{H-eq}\\
{\large{\Box}}\phi+\frac{1}{6}e^{-2\phi}{H}_{\alpha\beta\gamma}
{H}^{\alpha\beta\gamma}
&=&0
\end{eqnarray}
where 
\begin{eqnarray}
^{(\phi)}{T}_{\mu\nu}&=&
\frac{1}{2}(\phi_{,\mu}\phi_{,\nu}
-\frac{1}{2}{g}_{\mu\nu}{g}^{\alpha\beta}
\phi_{,\alpha}\phi_{,\beta})\label{tphi}\\
^{(H)}{T}_{\mu\nu}&=&
\frac{1}{12}e^{-2\phi}\left(
3{H}_{\mu\lambda\kappa}{H}_{\nu}^{\;\;\lambda\kappa}
-\frac{1}{2}{g}_{\mu\nu}{H}_{\alpha\beta\gamma}
{H}^{\alpha\beta\gamma}\right)
\label{tH}
\end{eqnarray}

In four dimensions the antisymmetric tensor field strength can
be written in terms of the pseudoscalar field, $b$, as follows
\begin{eqnarray}
H^{\mu\nu\lambda}=e^{2\phi}\epsilon^{\rho\mu\nu\lambda}
b_{,\rho},\label{H1}
\end{eqnarray}
 and since the solutions including the axion $b$ can be obtained from pure
dilaton solutions via a $SL(2,I\!\!R)$ transformation  leaving
invariant the Einstein frame metric, we  will ignore this field
and concentrate on gravi-dilaton system in what follows.

In \cite{buon} the following two assumptions are required to hold
for an asymptotically past trivial (APT) initial state:
\begin{itemize}{}
\item[-] 
APT$_{1}$:  The string theory is weakly coupled, i.e. $g=e^{\phi/2}\ll 1.$
\item[-] 
APT$_{2}$: The curvature in string units is small.
\end{itemize}
APT$_{1}$ ensures that classical string theory is valid and string loop
corrections to (\ref{act}) can be ignored, whereas APT$_{2}$ means that $\alpha'$ corrections
to the same action are negligible.
Our starting point  will be to assume that in the asymptotic past the
universe is in a trivial state characterized by gravitational waves propagating in a flat 
space-time. Eventually, these plane waves will collide giving rise to non-trivial
geometries in the interaction region, and in particular, possibly, to the nucleation of 
PBB universes. Since plane waves are exact string vacua, APT$_{2}$ is automatically 
satisfied in the space-time regions before the interaction \cite{horstief}. However, 
it is not at all clear that the exact conformal invariance of the background also holds for 
all the possible solutions describing 
the interaction region. From physical considerations one would expect 
to have at least one of these solutions in the interaction region corresponding to an exact 
string background, and that this solution would smoothly match the incoming plane waves
along the null boundaries of the interaction region (see below).

The space-times representing interactions of plane  
waves have two commuting Killing vectors $\zeta_{1}$, $\zeta_{2}$ and it is possible 
to chose a system of adapted coordinates $(u,v,x,y)$ in which $\zeta_{1}=\partial_{x}$ 
and $\zeta_{2}=\partial_{y}$, whereas the longitudinal coordinates $(u,v)$ are
null. In the $u$-$v$ plane, the resulting space-time can be divided into four 
different regions (cf. Fig. 1) \cite{kp,szek,grif}:
\begin{itemize}
\item[-]
Region I ($u<0$, $v<0$) is flat space-time, described by the usual
Minkowski line element 
$$
ds^2_{I}=-2du dv+dx^2+dy^2
$$
and a constant dilaton.
\item[-]
Region II ($u\geq 0$, $v\leq 0$) is incoming plane wave 1,
described by 
\begin{eqnarray}
ds^2_{II}=-2du dv+F_{1}^{2}(u)dx^2+G_{1}^{2}(u)dy^2,
\label{one}
\end{eqnarray}
and dilaton field ${\phi_{1}(u)}$.
\item[-]
Region III ($u\leq 0$, $v\geq 0$) is incoming plane wave 2,
described by 
\begin{eqnarray}
ds^2_{III}=-2du dv+F_{2}^{2}(v)dx^2+G_{2}^{2}(v)dy^2,
\label{two}
\end{eqnarray}
along with ${\phi_{2}(v)}$.
\item[-]
Region IV  ($u\geq 0$, $v\geq 0$) is the interaction region,
described by
\begin{eqnarray}
ds^2_{IV}=-2e^{-F}du dv+G
(e^{\psi}dx^2+e^{-\psi}dy^2),
\label{intreg}
\end{eqnarray}
where $F(u,v)$, $G(u,v)$ and $\psi(u,v)$, as well as the dilaton field $\phi(u,v)$, 
are functions of both $u$ and $v$.

\end{itemize}

\begin{figure}
\centerline{\epsfig{file=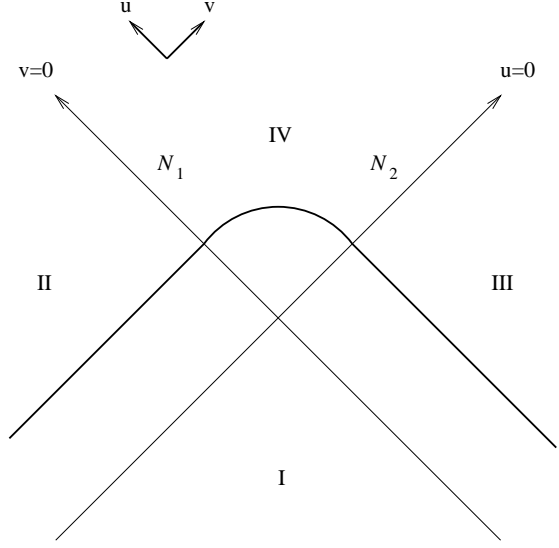, width=2.9in}}
\caption{Region I is the flat background space-time, region II and III describe
the approaching plane waves and region IV is the interaction region.}
\end{figure}

 In this problem the initial data are most conveniently  posed on the null surfaces 
$N_{1}\cup N_{2}$, where 
$N_{1}\equiv\{v=0,u\geq0\}$ and
$N_{2}\equiv\{u=0,v\geq0\}$, which is the
boundary of the interaction region IV.
In the interior of this region one of  Einstein's equations reads
\begin{eqnarray}
G_{uv}=0
\end{eqnarray}
which is solved by \cite{szek,FI,grif}
$$
G=a(u)+b(v)=1-(\alpha u)^{n}-(\beta v)^{m}.
$$
Here $\alpha^{-1}$ and $\beta^{-1}$ are arbitrary positive length scales
fixing the focal lengths of the incoming waves that we will set to 1 in
the following. On the other hand, the integers $n$ and $m$ are determined by the
boundary conditions.

Introducing two new coordinates $\xi$ and $z$ defined by ($\alpha=\beta=1$)
\begin{eqnarray}
\xi&\equiv & a(u)+b(v)=1-u^{n}-v^{m} \label{xin}\\
z&\equiv & a(u)-b(v)=u^{n}-v^{m} \label{zm}
\end{eqnarray}
the metric (\ref{intreg}) in the interaction region takes the
familiar Einstein-Rosen form
\begin{eqnarray}
ds^2=e^{f}(-d\xi^2+dz^2)+\xi(e^{\psi}dx^2+e^{-\psi}dy^2).
\end{eqnarray}
The equations of motion for the metric functions and the dilaton are given by
\begin{eqnarray}
\ddot{\psi}&+&\frac{1}{\xi}\dot{\psi}-\psi''=0\\
\ddot{\phi}&+&\frac{1}{\xi}\dot{\phi}-\phi''=0\\
\dot{f}&=&-\frac{1}{2\xi}+\frac{\xi}{2}
\left(\dot{\psi}^2+\psi'^2\right)
+\frac{\xi}{2}
\left(\dot{\phi}^2+\phi'^2\right)
\label{fdot}\\
f'&=&\xi\dot{\psi}\psi'+\xi\dot{\phi}\phi'
\label{fprime}
\end{eqnarray}
where  a dot and a prime  denote differentiation with respect to $\xi$ and $z$ respectively.
The equations for $\psi$ and $\phi$ can be solved in terms
of Bessel and Neumann functions by \cite{FI}
\begin{eqnarray}
V=k\ln\xi+{\cal L}\{A_{\omega}\cos[\omega(z+z_0)]
J_0(\omega\xi)\}
+{\cal L}\{B_{\omega}\cos[\omega(z+z_0)]
N_0(\omega\xi)\}\, .
\label{gs}
\end{eqnarray}
Here $V$ stands for either $\psi$ or $\phi$ and ${\cal L}\{...\}$
denotes arbitrary linear combinations of the terms
in curly brackets including those of the
form $\int_0^{\infty} A_{\omega}\cos[\omega(z+z_0)]
J_0(\omega\xi)$, $\int_0^{\infty} B_{\omega}\cos[\omega(z+z_0)]
N_0(\omega\xi)$.

In order to relate the asymptotic behaviour near the singularity
at $\xi=0$ to the initial data given on the boundary
of the interaction region at $\{(u,0)\}\cup\{(0,v)\}$ it
is useful to introduce yet another set of coordinates
$r$ and $s$ defined by
$$
r\equiv \xi-z,\hspace{1cm}
s\equiv \xi+z.
$$
In this case the equations for $\psi$ and $\phi$
take the form
\begin{eqnarray}
\psi_{,rs}+\frac{1}{2(r+s)}(\psi_{,r}+\psi_{,s})&=&0\\
\phi_{,rs}+\frac{1}{2(r+s)}(\phi_{,r}+\phi_{,s})&=&0.
\label{pwe}
\end{eqnarray}

These two equations, together with the initial data on the null boundaries
of the interaction region $\{\psi(r,1),\psi(1,s)\}$
and $\{\phi(r,1),\phi(1,s)\}$
pose a well defined initial value problem. Both 
$\psi(r,s)$ and $\phi(r,s)$ are $C^{1}$ (and
piecewise $C^{2}$) functions. 
This problem was first solved by Szekeres \cite{szek} in the case of
pure gravitational waves.
Here the notation of Yurtsever \cite{yurt} is used,
\begin{eqnarray}
\psi(r,s)&=&\int_{1}^{s}ds'
\left[\psi_{,s'}(1,s')+\frac{\psi(1,s')}{2(1+s')}\right]
\left[\frac{1+s'}{r+s}\right]^{\frac{1}{2}}
{\cal P}_{-\frac{1}{2}}
\left[1+2\frac{(1-r)(s'-s)}{(1+s')(r+s)}\right]
\nonumber\\
& &
+\int_{1}^{r}dr'
\left[\psi_{,r'}(r',1)+\frac{\psi(r',1)}{2(1+r')}\right]
\left[\frac{1+r'}{r+s}\right]^{\frac{1}{2}}
{\cal P}_{-\frac{1}{2}}
\left[1+2\frac{(1-s)(r'-r)}{(1+r')(r+s)}\right]
\label{pgs}\\
\phi(r,s)&=&\int_{1}^{s}ds'
\left[\phi_{,s'}(1,s')+\frac{\phi(1,s')}{2(1+s')}\right]
\left[\frac{1+s'}{r+s}\right]^{\frac{1}{2}}
{\cal P}_{-\frac{1}{2}}
\left[1+2\frac{(1-r)(s'-s)}{(1+s')(r+s)}\right]
\nonumber\\
& &
+\int_{1}^{r}dr'
\left[\phi_{,r'}(r',1)+\frac{\phi(r',1)}{2(1+r')}\right]
\left[\frac{1+r'}{r+s}\right]^{\frac{1}{2}}
{\cal P}_{-\frac{1}{2}}
\left[1+2\frac{(1-s)(r'-r)}{(1+r')(r+s)}\right]
\label{phigs}
\end{eqnarray}
where ${\cal P}_{-\frac{1}{2}}(x)$ is a Legendre function.
It is  important to stress here that, in order to study the
behaviour near the singularity, we are only concerned with the 
``nonzero mode'' of the dilaton. Therefore, $\phi(r,s)$ in
Eq. (\ref{phigs}) is normalized in such a way that $\phi(1,1)=0$. However, 
we can always add an arbitrary constant to the dilaton field and
still have a solution to the wave equation (\ref{pwe}). In
particular,  we can tune the string coupling constant
to small values in regions II and III (and of course in I as well)
without affecting the structure of the singularity in region IV.

Near the singularity at $\xi=0$ Kasner behaviour
is expected, and it is known that space-times admitting two
abelian space-like Killing vectors with parallel polarization
have an asymptotically velocity dominated singularity
\cite{mon}  so that  curvature effects become negligible
there.

The nice feature of the colliding plane wave space-times is
that the initial value problem is well posed and can
be solved exactly. This allows to relate the Kasner exponents
which describe the behaviour of the metric near the
singularity to the initial data given on the null boundaries
of the interaction region.
In order to find this relationship we expand the Bessel functions
around $\xi=0$, and proceeding along the lines of  Yurtsever's work \cite{yurt} write the
following decomposition
\begin{eqnarray}
\psi(\xi,z)&=&\epsilon(z)\ln\xi+d(z)+H(\xi,z),\nonumber\\
\phi(\xi,z)&=&\varphi(z)\ln\xi+\tilde{d}(z)
+\tilde{H}(\xi,z),\nonumber
\end{eqnarray}
where $\epsilon(z)$, $\varphi(z)$, $d(z)$ and $\tilde{d}(z)$ are
independent of $\xi$, and $H(\xi,z)$, $\tilde{H}(\xi,z)$ vanish in
the limit $\xi\rightarrow 0$ [or $r+s\rightarrow 0$ in $(r,s)$ coordinates]. 
Thus, the Kasner exponents in this limit are
determined entirely by the coefficients of the logarithmic terms in the 
above expansions. 

These coefficients can be computed directly from Eqs. (\ref{pgs}) and (\ref{phigs}). For
$\epsilon(z)$ one finds 
\begin{eqnarray}
\epsilon(z)&=&\frac{1}{\pi\sqrt{1+z}}
\int_{z}^{1}ds \left[(1+s)^{\frac{1}{2}}\psi(1,s)
\right]_{,s}
\left(\frac{s+1}{s-z}\right)^{\frac{1}{2}}\nonumber\\
&+&
\frac{1}{\pi\sqrt{1-z}}
\int_{-z}^{1}dr \left[(1+r)^{\frac{1}{2}}\psi(r,1)
\right]_{,r}
\left(\frac{r+1}{r+z}\right)^{\frac{1}{2}},
\label{ep}
\end{eqnarray}
and a similar expression holds for the dilaton
\begin{eqnarray}
\varphi(z)&=&\frac{1}{\pi\sqrt{1+z}}
\int_{z}^{1}ds \left[(1+s)^{\frac{1}{2}}\phi(1,s)
\right]_{,s}
\left(\frac{s+1}{s-z}\right)^{\frac{1}{2}}\nonumber\\
&+&
\frac{1}{\pi\sqrt{1-z}}
\int_{-z}^{1}dr \left[(1+r)^{\frac{1}{2}}\phi(r,1)
\right]_{,r}
\left(\frac{r+1}{r+z}\right)^{\frac{1}{2}}.\label{varp}
\end{eqnarray}

By introducing the leading logarithmic behaviour of both $\psi(\xi,z)$ and $\phi(\xi,z)$ into
the equations for $f(\xi,z)$, Eqs. (\ref{fdot}) and (\ref{fprime}), we readily get the
solution for the metric function $f(\xi,z)$ near the singularity at $\xi=0$ to be
\begin{eqnarray}
f(\xi,z)\simeq\frac{1}{2}[\epsilon^{2}(z)+
\varphi^{2}(z)-1]\ln\xi.
\end{eqnarray}
Hence the asymptotic behaviour of the metric when $\xi\rightarrow 0$ is
given by
\begin{eqnarray}
ds^2=\xi^{a(z)}(-d\xi^2+dz^2)+\xi^{1+\epsilon(z)}dx^2
+\xi^{1-\epsilon(z)}dy^2
\label{m2}
\end{eqnarray}
where $a(z)\equiv\frac{1}{2}[\epsilon^{2}(z)+ \varphi^{2}(z)-1]$
(cf. also \cite{in_pbb}). 

Thus, we have 
completely specified the asymptotic behaviour of the metric near the caustic singularity
in terms of the initial data for $\psi(\xi,z)$ and $\phi(\xi,z)$, as encoded by the
functions $\epsilon(z)$ and $\varphi(z)$. In the following section we will use this
result to study the initial conditions in the collision problem leading to PBB inflation.

\section{Conditions for pre-big-bang inflation}

Now that the relation between the asymptotic form of the metric (\ref{m2}) 
and the initial conditions on the boundary of the interaction region is given, we
can address the problem of determining what kind of initial data lead to PBB 
inflationary solutions. Transforming the solution (\ref{m2})
to the string frame and switching,
 once in the string frame, from conformal   to cosmic time 
 we find the following Kasner exponents (generically, these
will be functions of $z$)
\begin{eqnarray}
p_{1}(z)&\equiv&\frac{1+\epsilon(z)+\varphi(z)}
{b(z)+2}\nonumber\\
p_{2}(z)&\equiv&\frac{1-\epsilon(z)+\varphi(z)}
{b(z)+2}\nonumber\\
p_{3}(z)&\equiv&\frac{b(z)}{b(z)+2}
\label{ke}
\end{eqnarray}
where $b(z)\equiv\frac{1}{2}[\epsilon^{2}(z)+
\varphi^{2}(z)+2\varphi(z)-1]$ and the subscripts 1,2,3 correspond 
to the $x,y,z$ directions respectively. These exponents
 satisfy the usual conditions for 
dilaton-vacuum Kasner solutions in the string frame 
$$
\sum_{i=1}^{3} p_{i}(z)=1+{2\varphi(z)\over b(z)+2}\, , \hspace{1cm}
\sum_{i=1}^{3} p_{i}(z)^{2}=1\, .
$$

The conditions for PBB inflation ($p_{1},p_{2},p_{3}<0$) are then 
translated into the following conditions on the functions $b(z)$, $\epsilon(z)$
and $\varphi(z)$:
\begin{eqnarray}
-2<b(z)<0\\
1-\epsilon(z)+\varphi(z)<0 \label{rect1}\\
1+\epsilon(z)+\varphi(z)<0 \label{rect2}\, .
\end{eqnarray}
Actually, the first inequality, when expressed in terms of $\epsilon(z)$ and
$\varphi(z)$, reads
\begin{eqnarray}
\epsilon(z)^{2}+[\varphi(z)+1]^{2}<2
\label{circle}
\end{eqnarray}
Thus, the set of points in the $\epsilon(z)$-$\varphi(z)$ plane for which we
get PBB inflationary solutions near the singularity is the quadrant of the 
circle defined by (\ref{circle}) and bounded by the lines (\ref{rect1}) and (\ref{rect2})
as shown in Fig. 2. This quadrant is inscribed on the square defined by
$|\epsilon(z)|<\sqrt{2}$ and $|\varphi(z)+1|<\sqrt{2}$.

\begin{figure}
\centerline{\epsfig{file=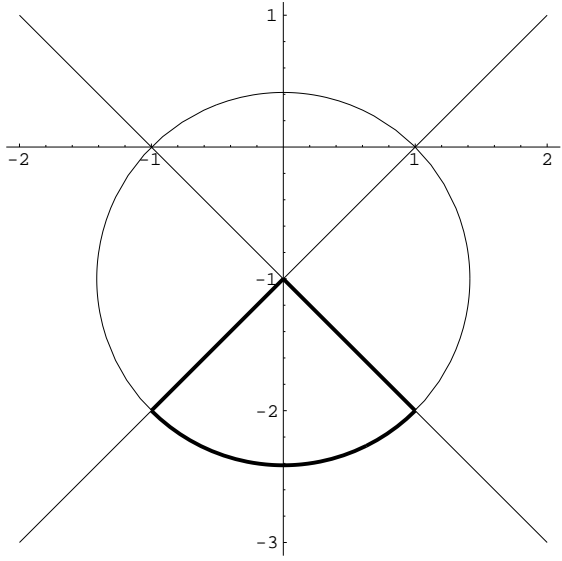, width=2.9in}}
\caption{Parameter space on the $\varphi(z)$ vs. $\epsilon(z)$ plane. The thick line
bounds the region representing those models for which inflation occurs in all directions.}
\end{figure}

Since for the inflating models $\varphi(z)<0$, 
$$
g_{\rm eff} \sim (-\xi)^{{1\over 2}\varphi(z)} \longrightarrow +\infty, \hspace{1cm}
(\xi\rightarrow 0^{-}) 
$$
so the effective string coupling constant diverges at the
singularity. As a consequence, near $\xi=0$ quantum corrections 
will be large and will lead, hopefully, to  regularization of the singularity and 
a graceful exit into the post-big-bang (i.e. FRW) phase.

Now the question of genericity of PBB inflation can be posed. In order to answer
this we impose APT on the initial data. APT$_{1}$ demands that the string coupling 
constant has to be small. Since our equations are invariant under the shift of the
dilaton by a constant $\phi\rightarrow \phi+{\rm constant}$ we can always fix 
this constant in such a way that the string theory is weakly coupled in regions I, II and III
and thus also at the null boundaries $N_{1}=\{v=0,u>0\}$, $N_{2}=\{u=0,v>0\}$.

On the other hand, APT$_{2}$ is implemented by requiring small curvatures. 
We can mathematically express this condition by demanding the
components of the Weyl tensor to be small on the initial null hypersurfaces 
$N_{1}$ and $N_{2}$. On $N_{1}$ the only non-vanishing component is $\Psi_{4}$ 
and it is given by
\begin{eqnarray}
\left.\Psi_{4}\right|_{N_{1}}=-\frac{1}{2}\psi_{uu}
+\frac{1}{4}\psi_{u}\left[2\frac{n-1}{u}+\frac{3n\, u^{n-1}}
{1-u^{n}}
-\frac{1- u^{n}}{n\, u^{n-1}}
(\psi_{u}^{2}+\phi_{u}^{2})\right]
\label{wsco}
\end{eqnarray}
where we have used the notation $\psi_{1}(u)\equiv \psi(u,v=0)$ and 
$\phi_{1}(u)\equiv\phi(u,v=0)$.
On the other hand, on $N_{2}$ only $\Psi_{0}$ is non-zero, and we find
\begin{eqnarray}
\left.\Psi_{0}\right|_{N_{2}}=-\frac{1}{2}\psi_{vv}+\frac{1}{4}\psi_{v}
\left[2\frac{m-1}{v}+\frac{3m\, v^{m-1}}{1-v^{m}}
-\frac{1-v^{m}}{m\, v^{m-1}}
(\psi_{v}^{2}+\phi_{v}^{2})\right] \label{wsct}
\end{eqnarray}
where now\footnote{In writing $\Psi_{4}$ and $\Psi_{0}$ we have set 
the original focal lengths of the incoming waves $\alpha^{-1}$, $\beta^{-1}$ to 1. 
We can restore these length scales by writing
$\Psi_{4}|_{N_{1}}\rightarrow \alpha^{-2} \Psi_{4}|_{N_{1}}$ and 
$\Psi_{0}|_{N_{2}}\rightarrow \beta^{-2} \Psi_{0}|_{N_{2}}$
in Eqs. (\ref{wsco}) and (\ref{wsct}) respectively. Since the string length $\ell_{st}$ is the
natural scale of the problem, we take $\alpha^{-1}=\beta^{-1}=\ell_{st}=1$ and
measure all curvatures in string units.} 
$\psi_{2}(v)\equiv \psi(u=0,v)$ and $\phi_{2}(v)\equiv\phi(u=0,v)$.

Thus, in order to satisfy generically the conditions of having low curvature in string 
units, $\Psi_{4}|_{N_{1}}, \Psi_{0}|_{N_{2}}\ll 1$,
we have to demand that all the derivatives of the metric functions $\psi_{1}(u)$, $\psi_{2}(v)$
that appear in expressions (\ref{wsco}) and (\ref{wsct}), as well as 
the corresponding derivatives of the dilaton field on the boundaries, $\phi_{1}(u)$, $\phi_{2}(v)$
are much smaller than $1$. From the condition $\psi(u=0,v=0)=0$ and $\phi(u=0,v=0)=0$ [in the latter case 
by $\phi(u,v)$ we denote just the ``nonzero mode'' of the dilaton] and the smallness of 
the derivatives
we conclude that the functions $\psi_{1}(u)$, $\psi_{2}(v)$ as well as $\phi_{1}(u)$ and 
$\phi_{2}(v)$ are approximately constant and close to zero. Consequently, the initial data compatible with 
APT will satisfy
\begin{eqnarray}
\psi(1,s)\simeq \mu_{1}={\rm constant}\ll 1, \hspace{1cm} 
\psi(r,1)\simeq \mu_{2}={\rm constant}\ll 1,
\label{pp}
\end{eqnarray}
and similar relation for the ``nonzero mode'' of the dilaton field
\begin{eqnarray}
\phi(1,s)\simeq \nu_{1}, \hspace{1cm} \phi(r,1) \simeq \nu_{2} 
\label{phph}
\end{eqnarray}
where again $\nu_{1}$ and $\nu_{2}$ are constants much smaller than 1.
If we now make use of the expressions (\ref{ep}) and (\ref{varp}) that give
us the gravitational and matter source functions $\epsilon(z)$ and $\varphi(z)$
in terms of the initial data, we find that $\epsilon(z)$ is given by 
\begin{eqnarray}
\epsilon(z)\simeq\left(\frac{1-z}{1+z}\right)^{\frac{1}{2}}
\frac{\mu_{1}}{\pi}+
\left(\frac{1+z}{1-z}\right)^{\frac{1}{2}}
\frac{\mu_{2}}{\pi}.
\end{eqnarray}
This expression implies that $\epsilon(z)\ll 1$ for a large range of
values of $z\in (-1,1)$, as long as $\mu_{1},\mu_{2}\ll 1$. 
On the other hand for the dilaton we get a similar relation
\begin{eqnarray}
\varphi(z)\simeq\left(\frac{1-z}{1+z}\right)^{\frac{1}{2}}
\frac{\nu_{1}}{\pi}+
\left(\frac{1+z}{1-z}\right)^{\frac{1}{2}}
\frac{\nu_{2}}{\pi}
\end{eqnarray}
and again, since $\nu_{1}, \nu_{2}\ll 1$, $\varphi(z)\ll 1$ for a large range of values of $z$.

Thus, we have found that, on general grounds, APT selects the values of 
$\epsilon(z)$-$\varphi(z)$ in a region around the origin $\epsilon(z)=\varphi(z)=0$.
If for $\epsilon(z)$ this is consistent with the coordinates of those points in 
parameter space corresponding to  the models for which the nucleation of PBB bubbles happens 
(see Fig. 2), in the case of $\varphi(z)$ the situation is not that good, since to achieve PBB
inflation we need $-1-\sqrt{2}<\varphi(z)<-1$. In any case, it is important to notice that the
bounds imposed by APT are not equally strong for $\psi(u,v)$ and $\phi(u,v)$. While in order to
fulfill APT$_{2}$ we need both the first and second derivatives of $\psi(u,v)$ to be much
smaller than 1 on $N_{1}$ and $N_{2}$, for $\phi(u,v)$ we need just to demand this same condition on
the {\it square} of the first derivative. Thus, APT is compatible with the
hierarchy between the constants $\mu_{1},\mu_{2}$ and $\nu_{1},\nu_{2}$. 

Nevertheless, the important conclusion we have reached is that, as a result of the gravitational wave 
collision, PBB inflation happens for a dense set of initial data, i.e. the PBB inflation becomes 
achievable in our scenario. This means that, once we have an inflationary solution, inflation 
is stable under small perturbations of the initial conditions that lead to small variations of the 
Kasner exponents.

\section{Particular solutions in the interaction region}

In the previous Section we have discussed the general
asymptotic behaviour of the solutions near the curvature 
singularity and have shown that one may completely specify the structure of the singularity
in terms of initial data posed on the null boundaries of the 
interaction region. In what follows, we discuss two particular examples of
metrics in this region. 

It is known that any metric with two commuting spatial Killing directions
describes the interaction region in a colliding wave problem provided 
the appropriate boundary conditions are met. Here we will 
concentrate our attention on two cases which we think are
 of physical relevance. First, we will study, in the light of our approach, the 
solution of Nappi and Witten which describe an inhomogeneous universe with closed
spatial sections of $S^{3}$ topology \cite{nappi}. The most interesting feature of 
this solution is the fact that it is an exact string background. After that
we will consider dilatonic generalizations of the Schwarzschild metric, in order to
make contact with the spherical collapse picture of Buonanno, Damour and 
Veneziano \cite{buon} and the PBB inflating Kantowski-Sachs universes \cite{fvm}.

\subsection{The Nappi-Witten cosmological solution}

The four-dimensional cosmological model studied by Nappi and Witten \cite{nappi} results 
as the target space theory of a $SL(2,I\!\!R)\times SU(2)/SO(1,1)\times U(1)$ gauged 
Wess-Zumino-Witten model. The solution contains, besides the metric, non-trivial values
for the dilaton and antisymmetric tensor field. Actually, the solution containing the 
non-vanishing B-field can be obtained by an 
$O(2,2;I\!\!R)$ rotation of the metric \cite{gp,gmv} (for a review see \cite{gpr})
\begin{eqnarray}
ds^{2}=-dt^{2}+dw^{2}+\tan^{2}{w}\, dx^{2}
+\cot^{2}{t} \, dy^{2},
\label{gpr}
\end{eqnarray}
together with the dilaton field
\begin{eqnarray}
\phi=\phi_{0}-\log (\sin^{2}{t}\,\cos^{2}{w}).
\label{dill}
\end{eqnarray}
The above line element may be thought of as a  product of two two-dimensional 
black holes with Euclidean and Lorentzian signatures, both being 
 exact string backgrounds \cite{witbh}, corresponding  to a
$SL(2,I\!\!R)/SO(1,1)\times SU(2)/U(1)$ coset model.

In the Einstein frame (\ref{gpr}) is given by 
\begin{eqnarray}
ds^{2}=e^{f(t,w)}(-dt^2+dw^2)+K(t,w)[e^{\psi(t,w)}dx^2+e^{-\psi(t,w)}dy^2]
\label{ER}
\end{eqnarray}
with
\begin{eqnarray}
f(t,w) &=& \log(1-\cos{2t})+\log(1+\cos{2w}) \nonumber \\
K(t,w) &=& \sin{2t}\,\sin{2w} \nonumber \\
\psi(t,w) &=& \log{\tan{t}}+\log{\tan{w}},
\end{eqnarray}
the dilaton field being given by (\ref{dill}).
To relate the Nappi-Witten solution with the plane 
 wave collision problem, it is convenient to switch to a 
different set of coordinates, namely 
\begin{eqnarray}
\xi=\sin{2t}\,\sin{2w}, \hspace{1cm} z=\cos{2t}\,\cos{2w}
\label{coorchn}
\end{eqnarray}
in which the solitonic nature\footnote{The name solitons is owed to
the inverse 
scattering technique used to obtain these solutions \cite{belin} rather than
to their physical properties. It was realised later
that in the diagonal case the gravi-soliton solutions
are related to the well known Lamb-Rosen pulses \cite{feinchar},
 see \cite{verd} for a review. Their role in the colliding wave problem was 
 discussed in \cite{FI}.} of the solutions can be made explicit.
 Bakas \cite{bak} also studied these solutions by applying 
the inverse scattering transform technique on a Kasner seed metric in a search
to relate the Geroch group to the symmetries of the string theory.
The above coordinate transformation can be inverted to give
\begin{eqnarray}
\log\tan{t}&=&{1\over 2}\left({\rm arc}\cosh{1+z\over \xi}-{\rm arc}\cosh{1-z\over \xi}\right),
\nonumber \\
\log\tan{w}&=&{1\over 2}\left({\rm arc}\cosh{1+z\over \xi}+{\rm arc}\cosh{1-z\over \xi}\right)
\end{eqnarray}
so in the new coordinates the metric function $\psi(\xi,z)$ is given by 
$$
\psi(\xi,z)= {\rm arc}\cosh\left({1+z\over \xi}\right)
$$
whereas the dilaton becomes
\begin{eqnarray}
\phi(\xi,z)={\rm arc}\cosh\left({1-z\over \xi}\right)-\log{\xi}.
\label{dilgpr}
\end{eqnarray}
In \cite{FI} 
it was pointed out  that the presence of at least two solitonic terms
provide  sufficient conditions for the continuity on the two different null boundaries 
of the region IV, if one is to interpret the spacetime in terms of plane wave interaction. 
This is indeed the case, since there is one soliton (the $\rm arc\cosh$ term) 
 associated with the 
dilaton solution and another one in the transverse metric function $\psi(\xi,z)$,
and the contribution to the boundary condition of each of those is equivalent as
if there where two solitons in the gravitational sector (for a general discussion
of the boundary conditions in the plane wave collision problem see Ref. \cite{grif},
in the string theory context see \cite{kehagias}) 

We  can now express both $\psi(\xi,z)$ and $\phi(\xi,z)$ in $(r,s)$-coordinates
($r=\xi-z$, $s=\xi+z$), so the initial data for our problem on the boundaries $N_{1}$ and
$N_{2}$ are specified by the functions
\begin{eqnarray}
\psi(1,s)=0, \hspace{1cm} \psi(r,1)=\log\left[{3-r+2\sqrt{2(1-r)}\over
1+r}\right]. 
\label{icpsi}
\end{eqnarray}
For the dilaton, on the other hand, we find
$$
\phi(1,s)=-2\log\left(1-\sqrt{1-s\over 2}\right),
\hspace{1cm}
\phi(r,1)=-\log\left({1+r\over 2}\right).
$$
By substituting these expressions into (\ref{ep}) and (\ref{varp}) we may directly study
the outcome of the collision near the singularity. We find that
$$
\epsilon(z)=-1, \hspace{1cm} \varphi(z)=-2.
$$
It can be easily seen that these values of $\epsilon(z)$ and $\varphi(z)$ lie just on the
boundary of the region of points for which the model undergoes PBB inflation.  
If we compute the Kasner exponents using (\ref{ke}) we find that 
$$
p_{1}=-1, \hspace{1cm} p_{2}=p_{3}=0,
$$
so there is only one inflating direction, while the other two  are ``frozen". 
Incidentally, the metric near the singularity corresponds to the T-dual of Milne space-time.

The original Nappi-Witten metric \cite{nappi} is obtained from the above solution  by the  
 $O(2,2;I\!\!R)$ rotation in string frame (followed by a rescaling of the $x$ coordinate, 
$x\rightarrow Bx$, see \cite{gpr}). Taking into account that 
$O(2,2;I\!\!R) \sim SL(2,I\!\!R)_{\tau}\times SL(2,I\!\!R)_{\rho}$ the required 
transformation can be written as
\begin{eqnarray}
\tau'=\tau, \hspace{1cm} \rho'={-1\over \rho+B}, \hspace{1cm} B\neq 0,
\label{ott}
\end{eqnarray}
where $\tau$ and $\rho$ are the usual K\"ahler and complex structure moduli constructed
from the string frame metric (see, for example, \cite{ajrd}). As discussed in \cite{ajrd}
the Einstein frame metric function $\psi(\xi,z)$ remains invariant under the 
$O(2,2;I\!\!R)$ rotation so we have
$$
\psi(\xi,z)_{\rm NW}={\rm arc}\cosh\left({1+z\over \xi}\right),
$$
while for the new dilaton we find
$$
\phi(\xi,z)_{\rm NW}=\phi(\xi,z)-\log\left(B^2+\xi^2 e^{2\phi(\xi,z)}\right)
$$
with $\phi(\xi,z)$ given by (\ref{dilgpr}). In addition, we have a non-vanishing 
value for the B-field
$$
B_{xy}(\xi,z)=-{B\over B^2+\xi^2 e^{2\phi(\xi,z)}}.
$$

Since $\psi(\xi,z)$ is left unchanged by the rotation, the initial conditions for 
$\psi(r,s)_{\rm NW}$ on the null boundaries $N_{1}$, $N_{2}$ are again given by (\ref{icpsi}).
Therefore, the gravitational source function $\epsilon(z)$ remains invariant.  
On the other hand, the dilaton does transform under (\ref{ott}), so the initial conditions for
the transformed dilaton are\footnote{Notice again that in giving the initial conditions for 
the dilaton we are restricting to the ``nonzero mode'' defined by $\phi(1,1)=0$ in 
$(r,s)$-coordinates. This means in particular that in the case of the Nappi-Witten 
solution we should  write the arbitrary additive constant $\phi_{0}$ in the dilaton as 
$\phi_{0}+\log{(1+B^{2})}$ in order to recover this nonzero mode when $\phi_{0}=0$.} 
\begin{eqnarray}
\phi(1,s)_{\rm NW}&=& -\log\left({1+r\over 2}\right), \\
\phi(r,1)_{\rm NW}&=& -\log\left[{3-s\over 2}+{1-B^2\over 1+B^2}\sqrt{2(1-s)}\right].
\end{eqnarray}
Using Eq. (\ref{varp}) we can check that the scalar source function $\varphi(z)_{\rm NW}$ vanishes.
Consequently, the model lies outside the inflationary region in the $\epsilon(z)$-$\varphi(z)$ plane.
If we evaluate the related Kasner exponents using (\ref{ke}) we get
$$
p_{2}=1, \hspace{1cm} p_{1}=p_{3}=0,
$$
so the metric asymptotically approaches the Milne regime as $\xi\rightarrow 0^{-}$.

In fact,  we may perform a  somewhat more general analysis; if we start with a solution 
characterized by some values of $\epsilon(z)$, $\varphi(z)$ within the inflationary region, 
after a generic $SL(2,I\!\!R)_{\rho}\subset O(2,2;I\!\!R)$ rotation the resulting metric 
near the singularity will be characterized by the new functions 
\begin{eqnarray}
\bar{\epsilon}(z)=\epsilon(z), \hspace{1cm} \bar{\varphi}(z) = -\varphi(z)-2.
\label{tr}
\end{eqnarray}
In particular, for every model leading to PBB inflation we have
$-\sqrt{2}-1<\varphi(z)<-1$, so the transformed function $\bar{\varphi}(z)$
will satisfy  $\bar{\varphi}(z)>-1$ and thus the metric
will not inflate at the singularity. Since transformations in the $SL(2,I\!\!R)_{\rho}$
factor of $O(2,2;I\!\!R)$ are the ones generating background values of the B-field,
one might be tempted to conclude that PBB inflation is not robust under the introduction
of this field. On the other hand, some of the models which were not inflating 
before the transformation was performed, may happen to inflate after.
 Incidentally, all points in Fig. 2 with $\varphi(z)\geq -1$  are preserved 
by transformations 
$$
\left(
\begin{array}{cc}
A & B \\
C & D 
\end{array}
\right) \in SL(2,I\!\!R)_{\rho}
$$
with $D\neq 0$, whereas they transform as in (\ref{tr}) when $D=0$.

\subsection{Dilatonic Schwarzschild-like metric}

In the original picture of Ref. \cite{buon}, the nucleation of PBB
bubbles comes through gravitational instability in the asymptotically trivial Universe. 
In our proposal, on the other hand, this nucleation is not so much due to gravitational
instability of the gravitational wave gas, but rather, the result of the mutual focusing of
these waves due to their nonlinear interaction. Needless to say that the initial conditions
are specified by quite different initial data in both pictures. What we propose here is 
the closest thing one might think to represent the decomposition of the initial data into 
plane waves in a non-linear theory. In what follows we will consider a set of initial data 
expressed as plane waves producing the same  behaviour in the interaction region, as if it were
a particular case of spherically symmetric gravitational collapse and will relate this to the 
solutions discussed in \cite{buon} and \cite{fvm}. The indication that this is possible relies on the 
previous studies \cite{gwbh} where the solution first obtained by Ferrari and
Ib\'{a}\~nez \cite{feri},
and representing part of the black hole, were investigated in detail.

To this end, we start again with the Gowdy metric (\ref{ER}) specifying
\begin{eqnarray}
f(t,w) &=& {1\over 2}[(a+1)^2+b^{2}-1]\log\sin{2t}-a\log(1+\cos{2t})  \label{sch1}\\
K(t,w) &=& \sin{2t}\sin{2w} \label{sch2}\\
\psi(t,w) &=& a\log{\tan{t}}+\log{\sin{2t}\sin{2w}}
\label{sch3}
\end{eqnarray}
and dilaton field
$$
\phi(t,w)=\phi_{0}+b\log{\tan{t}}
$$
where the two constants $a$ and $b$ satisfy the condition $a^{2}+b^{2}=4$. A common 
feature of this uniparametric family of solutions 
is that they are spatially homogeneous and of Kantowski-Sachs 
type with positive spatial curvature\footnote{This family of solutions corresponds 
to the family of 
closed Kantowski-Sachs cosmologies studied in \cite{fvm}, as can be seen by writing them
in the coordinate system $\tau={1\over 2}(1+\cos{2t})$, ${\rm x}=y$, $\varphi=\pi+2x$ and
$\theta=\pi+2w$.}. In particular, for $a=2$ and $b=0$, we obtain the ``inside-horizon region" 
of a
Schwarzschild black hole with $4M^2=1$. From (\ref{sch1})-(\ref{sch3}) we see that
the metrics are singular at $t=0 ,{\pi\over 2}$. Whenever $b\neq 0$ these are true 
curvature singularities with the curvature invariants blowing up. On the other hand, when $b=0$
($a=\pm 2$) the apparent singularity at $t=0$ is just a coordinate singularity, while the one
at $t={\pi\over 2}$ remains a curvature singularity.

We can now rewrite these solutions using $(\xi,z)$-coordinates defined by Eq. (\ref{coorchn}).
Doing so the metric function $\psi(\xi,z)$ and the dilaton field $\phi(\xi,z)$ are 
\begin{eqnarray}
\psi(\xi,z) &=& {a\over 2}\left({\rm arc}\cosh{1+z\over \xi}+{\rm arc}\cosh{1-z\over \xi}\right)
+\log\xi \nonumber \\
\phi(\xi,z) &=& {b\over 2}\left({\rm arc}\cosh{1+z\over \xi}+{\rm arc}\cosh{1-z\over \xi}\right)
\nonumber 
\end{eqnarray}
Changing into $(r,s)$-coordinates we readily get the initial conditions for
$\psi(r,s)$ on the null boundaries $N_{1}$, $N_{2}$
\begin{eqnarray}
\psi(1,s)&=&{a\over 2}\log\left[{3-s+2\sqrt{2(1-s)}\over 1+s}\right]+\log\left({1+s
\over 2}\right), \nonumber \\ 
\psi(r,1)&=&{a\over 2}\log\left[{3-r+2\sqrt{2(1-r)}\over 1+r}\right]+\log\left({1+r
\over 2}\right).
\nonumber
\end{eqnarray}
For the dilaton field we find
$$
\phi(1,s)={b\over 2}\log\left[{3-s+2\sqrt{2(1-s)}\over 1+s}\right], \hspace{1cm} 
\phi(r,1)={b\over 2}\log\left[{3-r+2\sqrt{2(1-r)}\over 1+r}\right].
$$
{}From these expressions we can  evaluate $\epsilon(z)$ and $\varphi(z)$ to get
$$
\epsilon(z)=1-a, \hspace{1cm} \varphi(z)=-b
$$
and since $a$ and $b$ satisfy $a^2+b^2=4$, we find that the values of $\epsilon(z)$ and $\varphi(z)$ 
lie on the circumference defined by
\begin{eqnarray}
[\epsilon(z)-1]^2+\varphi(z)^2=4.
\label{circ}
\end{eqnarray}
In Fig. 3 we have plotted this curve in the $\epsilon(z)$-$\varphi(z)$ plane. We find that it 
crosses the region of points for which there is PBB inflation as $\xi\rightarrow 0$. Actually, 
the points of the circumference (\ref{circ}) within the inflationary region correspond to the set of
models studied in \cite{fvm} for which both scale factors inflate (see Fig. 2 of Ref. \cite{fvm}).

\begin{figure}
\centerline{\epsfig{file=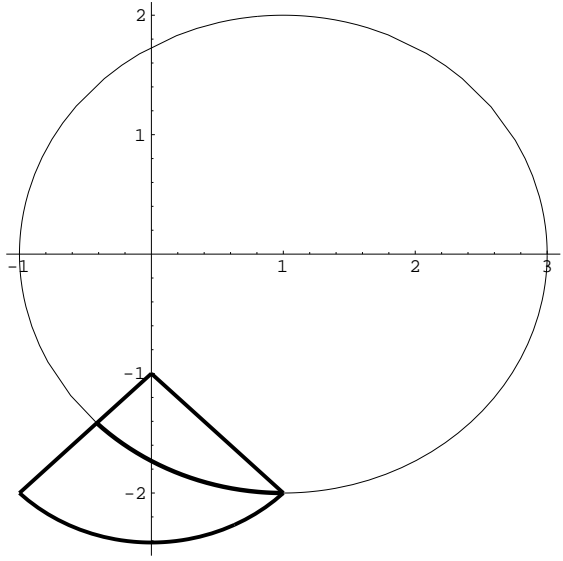, width=2.9in}}
\caption{Points in the $\varphi(z)$ vs. $\epsilon(z)$ plane corresponding to the 
family of Kantowski-Sachs models. The thicker sector of the circumference represents those 
models for which PBB inflation occurs.}
\end{figure}

We have focused our attention above to the family of deformations of the Schwarzschild black hole
labeled by a single parameter and in 
which homogeneity is preserved, i.e. the metric is of Kantowski-Sachs type. 
One may construct
more general deformations of the Schwarzschild metric by considering higher-dimensional moduli spaces.
Moreover, the family of Kantowski-Sachs solutions studied here can be extended to a 
two-parametric class of solutions with ``homogeneous'' longitudinal part of the
metric, defined by  
\begin{eqnarray}
f(t,w)&=&{1\over 2}[(a_{1}+a_{3})^{2}+(b_{1}+b_{3})^{2}-1]\log\sin{2t}-(a_{1}a_{3}+b_{1}b_{3})
\log\sin(1+\cos{2t}) \nonumber \\
K(t,w) &=& \sin{2t}\sin{2w} \nonumber \\
\psi(t,w)&=& a_{1}\log\tan{t}+a_{2}\log{\tan{w}}+a_{3}\log(\sin{2t}\sin{2w})
\nonumber
\end{eqnarray}
with the dilaton field
$$
\phi(t,w)=\phi_{0}+b_{1}\log\tan{t}+b_{2}\log{\tan{w}}+b_{3}\log{(\sin{2t}\sin{2w})}
$$
and the set of constants $\{a_{1},a_{2},a_{3}\}$ and $\{b_{1},b_{2},b_{3}\}$ satisfying the four 
conditions
\begin{eqnarray}
(a_{1}+a_{2})^2+(b_{1}+b_{2})^2 &=& 4 \nonumber \\
(a_{1}-a_{2})^2+(b_{1}-b_{2})^2 &=& 4 \nonumber \\
(a_{2}+a_{3})^2+(b_{2}+b_{3})^2 &=& 1 \nonumber \\
a_{2}a_{3}+b_{2}b_{3} &=& 0. \nonumber 
\end{eqnarray}
By solving these equations and studying the behaviour of the solution close to the singularity, 
we find that the resulting two-parametric family of models covers the 
region of the $\epsilon(z)$-$\varphi(z)$ plane defined by
$$
1\leq \epsilon(z)^2+\varphi(z)^2 \leq 9
$$
which indeed contains the set of points for which  PBB inflation occurs. However, if we want to 
relate this with the gravitational collapse picture of Ref. \cite{buon}  the solutions must 
possess a rotational symmetry and the only models in the family with an 
$SO(3)$ isometry are those with $a_{2}=b_{2}=b_{3}=0$, $a_{3}=1$, which  
precisely fall into the family of Kantowski-Sachs metrics that we have studied above.

\section{Conclusions and outlook}
In this  paper we have proposed a picture where the 
PBB inflation is realized starting from a trivial asymptotic state.
The idea is to start with  strictly plane waves moving in different directions which
interact  gravitationally at some stage to produce a space-time
singularity. 
The structure of the solutions close to
the singularity is of the Kasner type, and we were able to relate
analytically the initial data on the wave fronts to the Kasner
exponents in the string frame. Although maybe the collapsing regions
studied here are not as generic as for example those studied in \cite{buon},
our picture has two basic advantages:
\begin{itemize}
\item[1.]
The initial background is exact from the point of view of string propagation.
\item[2.]
One can completely determine the structure of the singularity in terms of the 
initial data provided by the incoming waves.
\end{itemize}
These two elements make of this scenario, at least, a solid test bench and a sort of 
theoretical laboratory for the PBB ideas.

We have seen that there exists a dense set of initial data leading to  
inflationary behaviour in the PBB phase and, therefore, there is a good chance 
for an inflationary universe to emerge during this phase.  Since the Kasner 
exponents actually carry a space dependence, i.e. they depend in general on one 
coordinate $z$, different regions with different Kasner exponents experience 
different types of inflation. Therefore, although we start with the collision 
of two plane waves we obtain near the singularity a rich structure
with the formation of different multiple PBB
inflationary bubbles. 

Although we have extracted these conclusions from 
a general analysis of the collision of two gravi-dilatonic waves, we have also
studied two concrete examples of initial conditions leading to different 
geometries in the interaction region. 
The first one leads to a Nappi-Witten solution in region IV and it has the
obvious interest of providing an exact string background also in this region.
The second example corresponds to a spherically symmetric interaction region 
that may describe gravitational collapse as the result of the collision.

Another motivation to look at plane wave space-times
as a probable initial state for the PBB universe may
come from thermodynamical considerations. One expects the
universe to start in the lowest possible state of gravitational
entropy. Let us suppose, arguing in the spirit of Penrose's hypothesis
\cite{penrose}, that we relate the gravitational entropy to some homogeneous function
constructed from all possible curvature invariants and that we normalize this function
to be vanishing when the invariants vanish. With such a function at hand we 
formulate a kind of ``Generalised Third Law of Thermodynamics'' and 
assign zero gravitational entropy to those spacetimes
for which {\em all} curvature invariants vanish. 
Interestingly, the FRW models do not fall into the class of zero entropy
models according to our definition  unlike in the Penrose's case. To justify
this, we argue  that the lowest entropy state, apart from being  the
simplest, must be an exact string background. The
space-times with all vanishing curvature invariants ({\it plane waves}) certainly do so, while
the FRW universe does not.

The additional support to consider the gravitational entropy content of
the plane wave geometry to be vanishing comes from yet a different, though
not unrelated argument. One would usually tend to relate the gravitational 
entropy with the phenomenon of quantum particle creation.
It is commonly accepted that quantum particle creation indicates whether
a system is endowed with nontrivial gravitational entropy.
Plane waves, due to their symmetry
and to the fact that all the curvature invariants vanish, do not
polarize the vacuum, so quantum particles are not created in the vicinity
of plane waves \cite{des}. This is consistent then with the hypothesis of 
assigning zero gravitation entropy to the plane wave.
Moreover, it looks as if time is not a player in the plane wave regime.
Due to the absence of the global Cauchy surface, one may consider such a pure 
plane wave geometry as ``timeless''. Therefore, until two such  waves interact, 
no notion of time as defined by entropy change is appreciated. What happens further 
is beyond the scope of this paper.

One of the most interesting issues, untouched here, is the problem of whether the
gravitational wave collision problem can be {\it globally} defined in terms of an exact
string background. Provided one starts with initial states that are exact string backgrounds, 
what are the conditions for the data to evolve into the interaction region without
breaking conformal invariance? We know that one such solution exists in this region, namely
the Nappi-Witten solution that we studied in Sec. 4.1. Therefore, since we start with plane
gravitational waves (which are exact string backgrounds) and we can make the transition over
the null boundaries as much differentiable as we like, the question remains of whether 
this implies conformal invariance in the interaction region. On purely physical grounds one
would be tempted to say that exact string backgrounds in the regions II and III of Fig. 1 will
smoothly (i.e. $C^{\infty}$) extend to an exact background in region IV, at 
least if the full string equations share the uniqueness properties of the Einstein equations. 

An interesting issue to address would be to refine the picture provided here to account for more 
realistic situations in which the primordial gravitational waves are
not plane but localized. It has been argued that for ``almost'' plane gravitational
waves singularities also occur as the result of their collision \cite{yuapw}. 
In the case of ``graviton beams'', there is also mutual focusing and maybe 
production of singularities \cite{vgb} that could serve as seeds for PBB bubbles.

Finally, there has been some work on the collision of
plane waves at Planckian energies \cite{planck} leading to black hole
nucleation through tunneling. This kind of scenarios might be a way to find a semi-classical
approximation of the formation of the singularity that could be
applied to the graceful exit problem in PBB cosmology. This,
and the thermodynamical ideas we have outlined above, we hope
to be able to discuss in the future.

\section*{Acknowledgements}
We are grateful to Jacob Bekenstein for  enlightening correspondence. Our
special thanks go to Gabriele Veneziano for his valuable comments on the
manuscript and interesting discussions.
A.F. acknowledges the support of University of the Basque Country Grant
UPV 122.310-EB150/98 and Spanish Science Ministry Grant PB96-0250.
K.E.K. is supported by the Swiss National Science Foundation.
The work of M.A.V.-M. has been supported by FOM ({\it Fundamenteel Onderzoek van de Materie}) 
Foundation and by University of the Basque Country Grants UPV 063.310-EB187/98 and UPV 
172.310-G02/99, and Spanish Science Ministry Grant AEN99-0315.
K.E.K. and M.A.V.-M. thank the Department of Theoretical Physics of 
The University of the Basque Country for hospitality.

\end{document}